\documentclass{article} % For LaTeX2e
\usepackage{graphicx}
\usepackage{amsmath}
\usepackage{amssymb}
\usepackage{xcolor}
\usepackage{hyperref}

\usepackage{iclr2023_conference,times}

\usepackage{amsmath,amsfonts,bm}

\def\eqref#1{equation~\ref{#1}}
\def\1{\bm{1}}

\DeclareMathAlphabet{\mathsfit}{\encodingdefault}{\sfdefault}{m}{sl}
\SetMathAlphabet{\mathsfit}{bold}{\encodingdefault}{\sfdefault}{bx}{n}

\usepackage{hyperref}
\usepackage{url}

\title{Improving Protein-peptide Interface Predictions in the Low Data Regime}

\author{Justin Diamond \& Markus Lill\\
Department of Pharmaceutical Science\\
Univiersity of Basel\\
Klingelbergstrasse 50, 4056 \\
Basel, Switzerland \\
\texttt{\{justin.diamond,markus.lill\}@unibas.ch} 
}

\iclrfinalcopy % Uncomment for camera-ready version, but NOT for submission.
\begin{document}

\maketitle

\begin{abstract}
We propose a novel approach for predicting protein-peptide interactions using
a bi-modal transformer architecture that learns an inter-facial joint distribution
of residual contacts. The current data sets for crystallized protein-peptide complexes are limited, making it difficult to accurately predict interactions between
proteins and peptides. To address this issue, we propose augmenting the existing data from PepBDB with pseudo protein-peptide complexes derived from the
PDB. The augmented data set acts as a method to transfer physics-based contextdependent intra-residue (within a domain) interactions to the inter-residual (between) domains. We show that the distributions of inter-facial residue-residue interactions share overlap with inter residue-residue interactions, enough to increase
predictive power of our bi-modal transformer architecture. In addition, this dataaugmentation allows us to leverage the vast amount of protein-only data available
in the PDB to train neural networks, in contrast to template-based modeling that
acts as a prior.

\end{abstract}

\section{Introduction}
   Compared to the experimental protein or protein small molecule structures, like those in PDB
(Berman et al., 2000), there is significantly less crystallized structures of protein-peptide complexes
which makes it difficult to build a model like AlphaFold in this setting. Given that amino acids
within and between domains share fundamental characteristics derive-able from Quantum Mechanics, there should be a transformation that improves the predictive power of protein-peptide interactions from purely intra-protein interactions. In principle, we want to determine the feasibility of
such transfer-ability methods by making simple augmentation of intra-protein residues to look more
like protein-peptide interactions. We take the augmentation as the cutting of a fake peptide sequence
from a protein sequence to obtain pseudo protein-peptide complexes derived from the PDB to mimic
that of the PepBDB (Huang et al., 2018). \\ \\
This could generalize to more heterogeneous contexts such as multiple peptides interactions or protein domain interfaces, making it more general and applicable to a wider range of problems. By
learning the coupling information where covalent bonding patterns are muteable, our model can
generalize to different sequences and even different types of molecules made up of amino acids, i.e.
branching or cyclic peptides. \\ \\
We frame our problem as classification via contact predictions of inter-facial residues of a peptide -
protein complex from sequence information alone, with the goal of predicting whether two residues
between a protein and peptide are within some Angstrom distance.

\section{Background}
	CASP (Critical Assessment of Techniques for Protein Structure Prediction) conference was initiated
in the early 90s to compare/develop the best computational techniques for in-silico prediction of
protein structure. In 2020s, they declared this problem generally solved with the advent of Deep
Learning techniques initially centered around template based models (Zhang et al., 2008), then
including RESNET architectures (Xu, 2017) from Computer Vision and finalized via a variation
of Large Language Models (LLM) seen in DeepMind’s AlphaFold (Senior et al., 2020) and Facebook’s Evolutionary Scale Models (Rives et al., 2019). The key take away was residue-residue
co-evolutionary couplings obtained from large Multiple Sequence Alignments (MSAs) contain
enough information to reconstruct 3D information of proteins with these new computational
techniques. There are generally two direct successor problems: crystal protein-protein/peptide
complex reconstruction and macromolecule dynamics i.e. computing equilibrium thermodynamic
state variables. We look to advance the first. \\ \\
Macromolecule 3D reconstruction from amino-acid sequences beyond proteins remains a key
goal. Structure reconstruction initially centered around transforming predicted residue-residue
contacts information into an energy potential which is used to optimized a protein structure.
DeepMind was the first to do this in a end-to-end differentiable manner. Most recently, diffusion
models (Wu et al., 2022) have shown great reconstruction abilities. \\ \\
We generalize the problem setting to obtaining inter residue-residue information from intra
residue-residue couplings.

\section{Methods}
\subsection{Bi-Modal Transformer}
The bi-modal transformer, similar to (Lu et al.,2016), takes in two inputs, the protein sequence $X_p$ of length $N$ and the peptide sequence $X_l$ of length $M$, both of which are tokenized into a sequence of amino acids.  The transformer uses the guided-attention mechanism to compute the
attention matrix between the protein and peptide sequences and at each layer, the protein and peptide
sequences are alternated such that at one layer the protein sequence embedding is updated with
respect to the peptide one, and the next vice versa.

In the first layer, the attention matrix, used to update the protein sequence, between the protein and peptide sequences is computed as follows:

$$A_{p,l} = \text{softmax}(\frac{Q_pK_l^T}{\sqrt{d_k}})V_p$$

Where $Q_p$, $K_l$, $V_p$ are the query, key, and value matrices for updating the protein sequence, and $d_k$ is the dimension of key used to scale the dot-product attention.

The hidden representation of the protein and peptide sequences are obtained by concatenating the attention matrices:

$$H_p = \text{concat}(A_{p,l})$$

In the next layer, the attention matrix between peptide and protein is computed:

$$A_{l,p} = \text{softmax}(\frac{Q_lK_p^T}{\sqrt{d_k}})V_l$$

Where $Q_l$, $K_p$, $V_l$ are the query, key, and value matrices for the peptide sequence.

$$H_l = \text{concat}(A_{l,p})$$

And this process continues for a fixed number of layers.

Finally, the objective function is obtained by performing logistic regression on the concatenated representation of the protein and peptide sequences:

$$Y = \sigma(W[H_p,H_l] + b)$$

Where $W$ is the weight matrix, $b$ is the bias term, and $\sigma$ is the sigmoid function.

The overall objective of the model is to maximize the likelihood of the correct interaction between the protein and peptide sequences, given by:

$$\mathcal{L} = \sum_{i=1}^{n}y_i \log p(y_i|x_i) + (1-y_i) \log (1-p(y_i|x_i))$$

where $n$ is the number of examples, $x_i$ is the pair of protein and peptide sequences, and $y_i$ is the true label indicating the interaction between the protein and peptide, which is defined as a 10 Angstrom cutoff threshold. 
\subsection{Evaluation Metrics}
To compare the true and learned distribution of contacts, we use the softmax function on the sum of the output of the model over all peptide residues for each protein residue to obtain a probability distribution over protein residues indicating the propensity of this residue to be in contact with one of the peptide residues. The softmax function is defined as:

$$P_i = \frac{e^{\sum_{j=1}^{M} x_{i,j}}}{\sum_{k=1}^{N}e^{\sum_{j=1}^{M} x_{k,j}}} $$

Where $P_i$ is the probability of protein residue $i$ interacting with any peptide residue, $x_{i,j}$ is the output of the model for the interaction between protein residue i and peptide residue j, N is the number of protein residues and M is the number of peptide residues.

Once we have the true and predicted probability distributions, we can use the Kullback-Leibler divergence (KL divergence) to evaluate the goodness of the model. The KL divergence between two probability distributions $P$ and $Q$ is defined as:

$$D_{KL}(P||Q) = \sum_{i=1}^{N} P_i \log\frac{P_i}{Q_i}$$

This measures the dissimilarity between the true and predicted probability distributions. The lower the KL divergence, the better the model is at reproducing the true distribution of contacts.
\subsection{Creating Augmented Protein-peptide Complexes }
We create the augmented protein-peptide dataset by obtaining protein structures from the Protein Data Bank (PDB). For each structure, we calculate the distance matrix between all pairs of amino acid residues:

$$D_{i,j} = ||R_i - R_j|| $$

Where $D_{i,j}$ is the distance between amino acid residues i and j, $R_i$ and $R_j$ are the coordinates of residues i and j respectively.

We then transform the distance matrix into a probability distribution over protein residues being in contact with other residues by summing over one dimension of the matrix and performing a softmax:

$$P_{i} = \frac{e^{\sum_{j=1}^{N} D_{i,j}}}{\sum_{k=1}^{N}e^{\sum_{j=1}^{N} D_{k,j}}} $$

Where $P_i$ is the probability of protein residue $i$, $D_{i,j}$ is the distance between amino acid residues $i$ and $j$, N is the number of amino acid residues.

We then sample indices from this distribution, where the indices indicate where the protein will be split. We also sample a uniform distribution from 10-50 amino acids in length to describe the length of the segment to be cut out from the protein. The protein is then concatenated together, with the cutout segment becoming the pseudo-peptide sequence. From the full distance matrix and the cutout segment, we can also calculate the pseudo-distance matrix.
\section{Results}

To evaluate the performance of our proposed method, we performed a series of experiments on a
dataset of protein-peptide complexes. Our goal was to assess the ability of our method to improve
contact predictions between the protein and peptide. \\ \\
In our first experiment, we predict contacts between residues of proteins and peptides determined
at a 10 Angstrom distance threshold. We then apply gaussian smoothing with convolutional kernels
with all the filter weights set to the value of 1, which has the effects of smoothing out the contact
signals as the predictions can be sensitive to small changes in residue positioning. \\ \\
As can be seen in Figure 1, the augmented data at times seems to help improve accuracy by moving
around contact groupings, sharpening the predictions, and refining the predictions. Although the
individual protein-peptide contact predictions are not always improved (there are cases where predictions are worse), from the test loss described below, in general, the prediction accuracy improves. \\ \\
\begin{figure}[htbp]
  \centering
  \begin{minipage}{0.3\textwidth}
    \includegraphics[width=\textwidth]{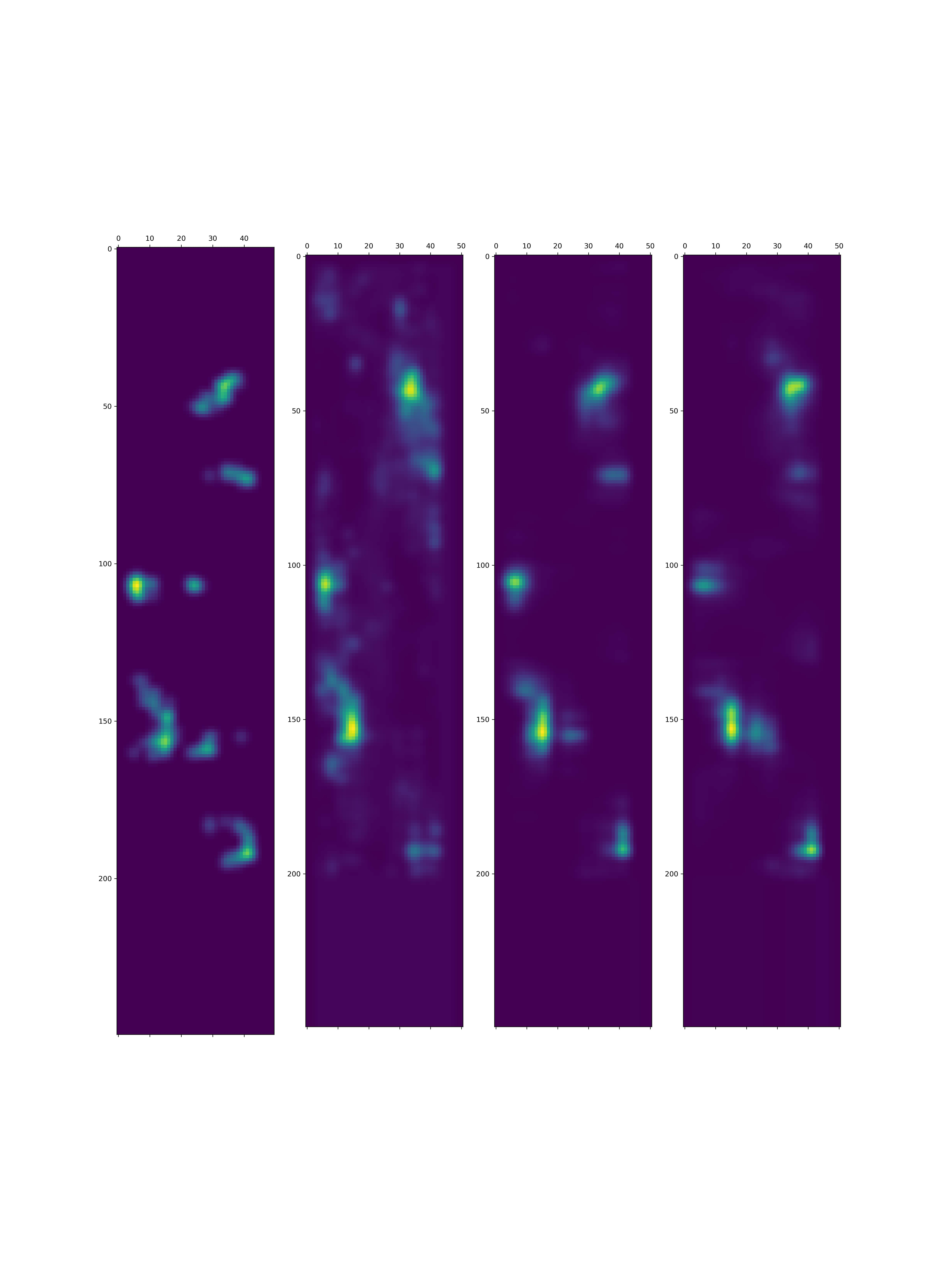}
  \end{minipage}
  \hfill
  \begin{minipage}{0.3\textwidth}
    \includegraphics[width=\textwidth]{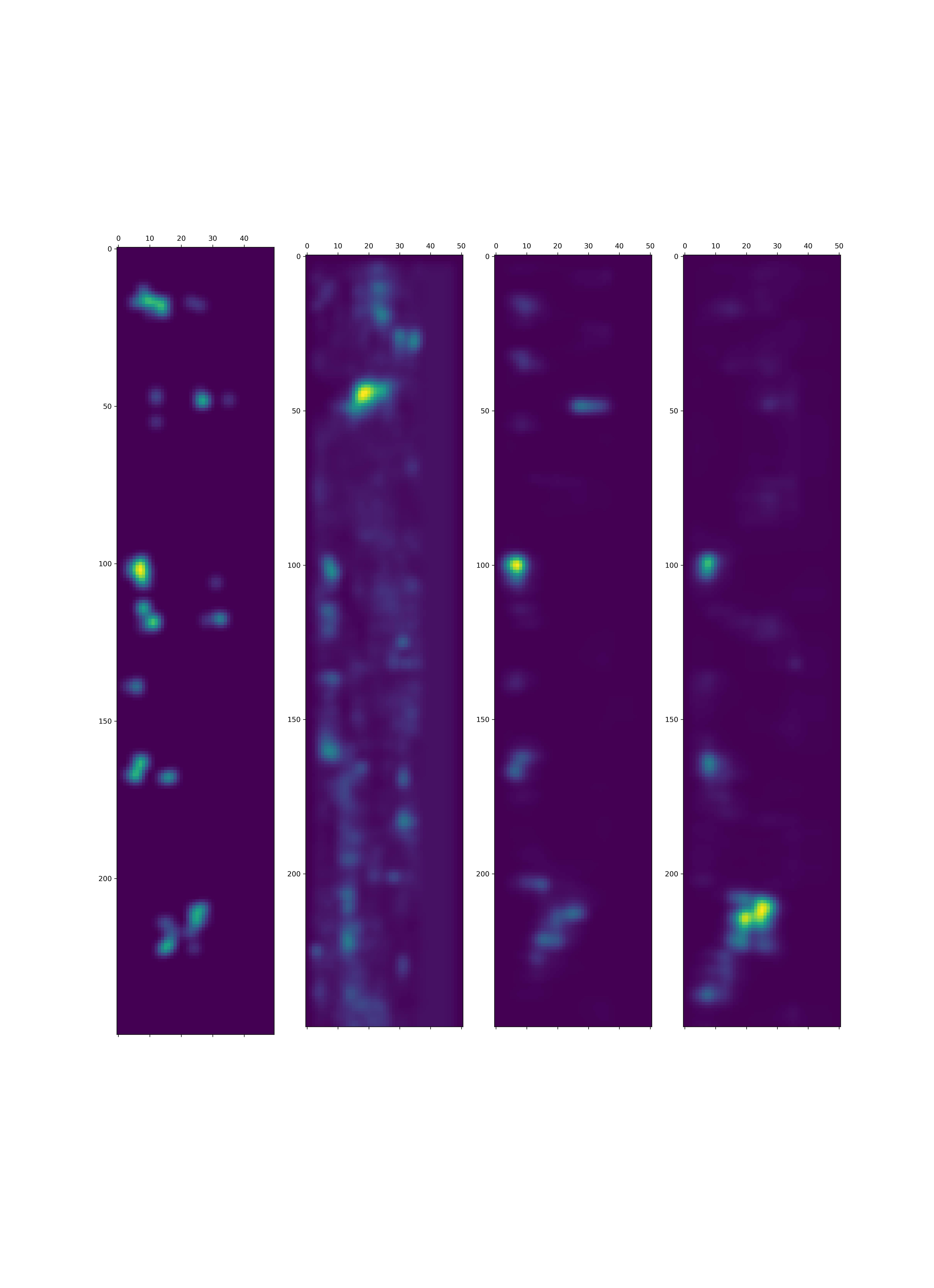}
  \end{minipage}
  \hfill
  \begin{minipage}{0.3\textwidth}
    \includegraphics[width=\textwidth]{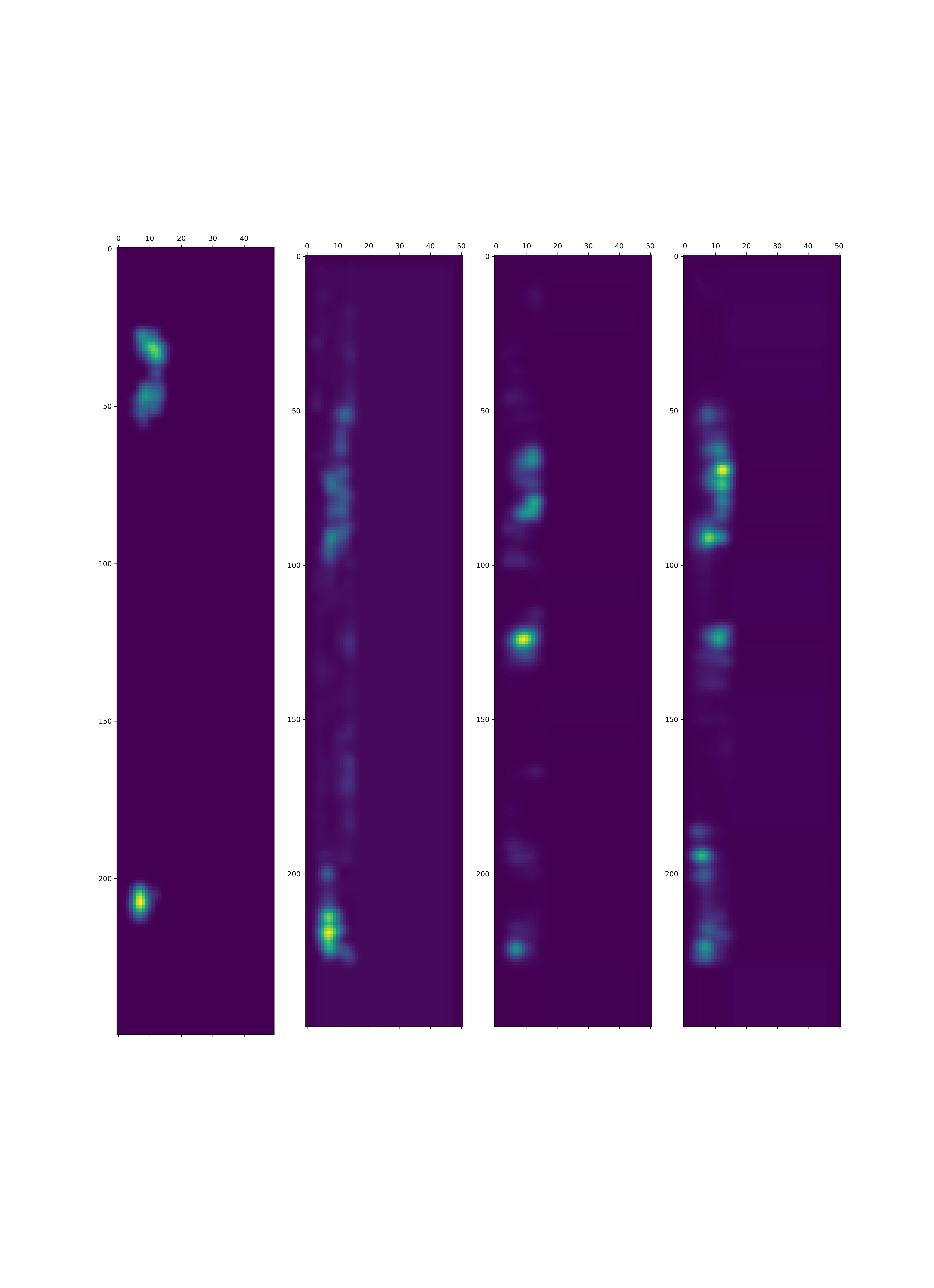}
    
  \end{minipage}
  
  \caption{Predicted contacts (from left to right) of True, no augmented data, 20000 augmented examples, and 30000 augmented examples. The X axis is ligand residue indices and the Y axis is protein residue indices. Brighter yellow regions indicate higher probabilities of a residue-residue contact between the peptide and protein.}
  \label{fig:example}
\end{figure}
Take for instance the far-right predictions in Figure 1. The second to the left of the four shows a less
resolved cluster of lower probability contacts towards the top compared to the augmented model.
This indicates that the augmented models increase their confidence further from the true contact
labels. \\ \\
In the middle four contact diagrams in contrast, the augmented models show higher probabilities
for some of the contact predictions which, retrospectively, accurately determines correct binding
regions. It also suggests that adding a certain amount of extra augmented training data can increase
accuracy further. However, certain precautions are necessary to make sure the model does not over
train on the most frequent set of augmented protein-peptide complexes. Adding more than 30000
augmented examples did not noticeably improve the test losses. \\ \\
In the second figure, we looked at the effects of data augmentation on test and validation loss. Some
complexes were similar to eachother, as measured by TMscore (Zhang et al., 2005), and if two
or more complexes are over a certain threshold, then they are made sure to be in either training,
validation, or test but not any other. This shows why the validation loss is slightly higher than the test loss. The augmented models outperform the baseline while the baseline admits a quick
downtrend in prediction accuracy due to the small size of the training set, while the augmented
models mediate this observation. \\ \\
\begin{figure}[htbp]
    \centering
    \includegraphics[width=0.7\textwidth]{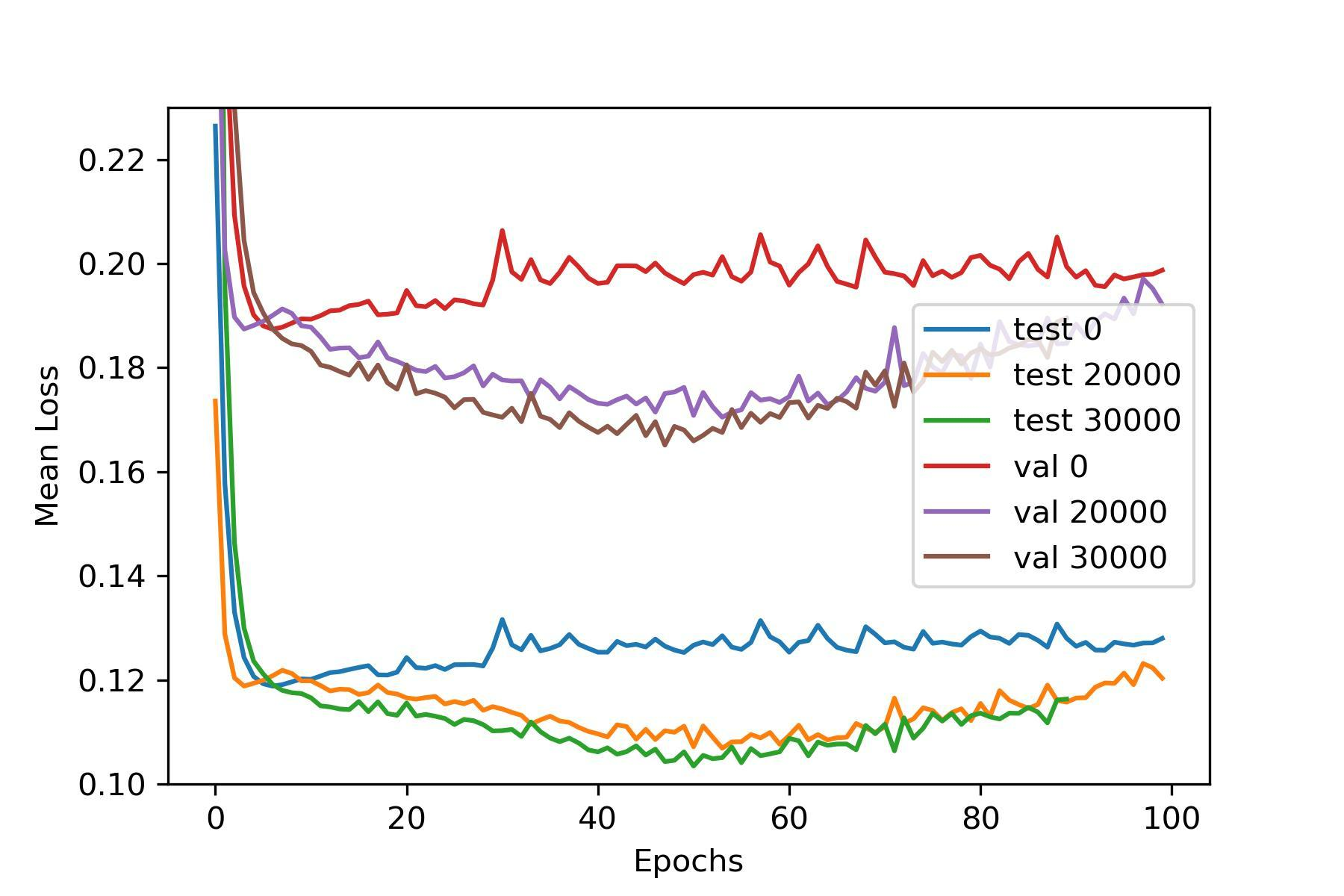}
    \caption{ Validation and test error for baseline (purple, orange), +20000 (red, blue), and +30000
(black, blue) augmented datasamples}
    \label{fig:your_label}
\end{figure}
In another experiment, we comparing the KL divergence of contact distributions on the protein or
peptide. As discussed above, to obtain distributions over the sequences of residues we apply sum
over the peptide (or protein ) sequences and apply softmax to obtain a probability distribution over
protein ( or peptide) sequences which are compared via the Kullback-Leibler divergence. This is
shown in Table 1 in the appendix, which shows improvements in binding site distribution predictions
with the augmented data. Lower KL divergence indicates that the augmented data could be useful
to a degree, but care must be taken to not weigh the augmented samples too heavily during training.
We also note that aggregating the contact probabilities to a distribution over peptide residues may
be useful in deciphering which residues are most important for binding.

\section{Discussion}
\subsection{Summary of the results}
Our results show that by augmenting a bi-modal transformer network with pseudo protein-peptide
complexes derived from the PDB, we are able to improve predictions of contacts between proteinpeptide complexes and their binding site predictions slightly. This is an important step towards
a better understanding of the physics of amino-acid interactions, and especially in the design of
transfer learning methods capable of generalizing across heterogeneous amino acid complexes.

\subsection{Future Research} Future directions could include adding Multiple Sequence Alignments to the model, which may help
improve sequence embeddings by implicitly encoding evolutionary couplings intra-sequence wise,
with the goal of obtaining evolutionary inter-residue couplings between sequences. \\ \\ 
Another direction would be to reconstruct protein-peptide complexes with these contact predictions
used as constraints to be satisfied. \\ \\
Lastly, it necessary to derive a similarity scheme between augmented examples so that the models
do not over train on the most common augmented residue-residue contacts, which would ideally
mean we need not weight the augmented datasets loss by a decreasing factor.

\section{Conclusion}
Our approach of augmenting a bi-modal transformer network with pseudo protein-peptide complexes represents a novel method for predicting protein-peptide interactions with limited data access.
It highlights the efficiencies of data-augmentation of transfer-learning. In principle, this approach
is a form of data distillation, where knowledge is transferred from a conditional distribution to generalized joint distribution and is particularly useful in cases where the distribution of the data is
unknown or difficult to obtain, and maximum likelihood methods are limited by the lack of data.

\section{References}
[1] Senior, A. W., Evans, R., Jumper, J., Kirshner, D., Korvink, J., Sylaidi, I., Kandel, A., Baran,
R., Ziebart, M., Frolov, D., et al. (2020). Improved protein structure prediction using potentials
from deep learning. Nature, 586(7830), 705–710. \\ 

[2] Zhang, Y. (2008). I-Tasser: a unified platform for automated protein structure and function
prediction. Nat. Methods, 5(2), 145–151. \\ 

[3] Xu, M., Yu, L., Song, Y., Shi, C., Ermon, S., Tang, J. (2022). GeoDiff: A Geometric Diffusion Model for Molecular Conformation Generation. In International Conference on Learning Representations (pp. 1-15). Retrieved from https://openreview.net/forumid=PzcvxEMzvQC \\ 

[4] Rives, A., Meier, J., Sercu, T., Goyal, S., Lin, Z., Liu, J., Guo, D., Ott, M., Zitnick, C. L., Ma,
J., Fergus, R. (2019). Biological Structure and Function Emerge from Scaling Unsupervised Learning to 250 Million Protein Sequences. PNAS. doi:10.1101/622803 \\ 

[5] Xu, J. (2017). Accurate De Novo Prediction of Protein Contact Map by Ultra-Deep Learning
Model. arXiv preprint arXiv:1703.06876. \\ 

[6] Lu, J., Yang, J., Batra, D., Parikh, D., Lee, D. D. (2016). Hierarchical Question-Image CoAttention for Visual Question Answering. arXiv preprint arXiv:1606.00061. \\ 

[7] Wen Z, He J, Tao H, Huang SY. PepBDB: a comprehensive structural database of biological
peptide-protein interactions. Bioinformatics. 2018 Jul 6. doi: 10.1093/bioinformatics/bty579. \\ 

[8] Berman, H. M., Westbrook, J., Feng, Z., Gilliland, G., Bhat, T. N., Weissig, H., Shindyalov,
I. N., and Bourne, P. E. (2000). The Protein Data Bank. Nucleic Acids Res. 28, 235-242. doi:
10.1093/nar/28.1.235. \\ 

[9] Wu, K. E., Yang, K. K., van den Berg, R., Zou, J. Y., Lu, A. X., Amini, A. P., Microsoft, and
Stanford. (2022, November 24). Protein structure generation via folding diffusion. [Online].
Available: https://arxiv.org/abs/2211.10297 \\ 

[10] Kolmogorov, A. (1998). On Tables of Random Numbers. Theoretical Computer Science.
207(2), 387–395. doi: 10.1016/S0304-3975(98)00075-9. MR 1643414. \\ 

[11] Zhang, Y. and Skolnick, J. (2005). TM-align: a protein structure alignment algorithm based on
the TM-score. Nucleic Acids Res. 33(7), 2302-2309. doi: 10.1093/nar/gki524.

\section{Appendix}
\begin{figure}[htbp]
  \centering
  \begin{minipage}{0.3\textwidth}
    \includegraphics[width=\textwidth]{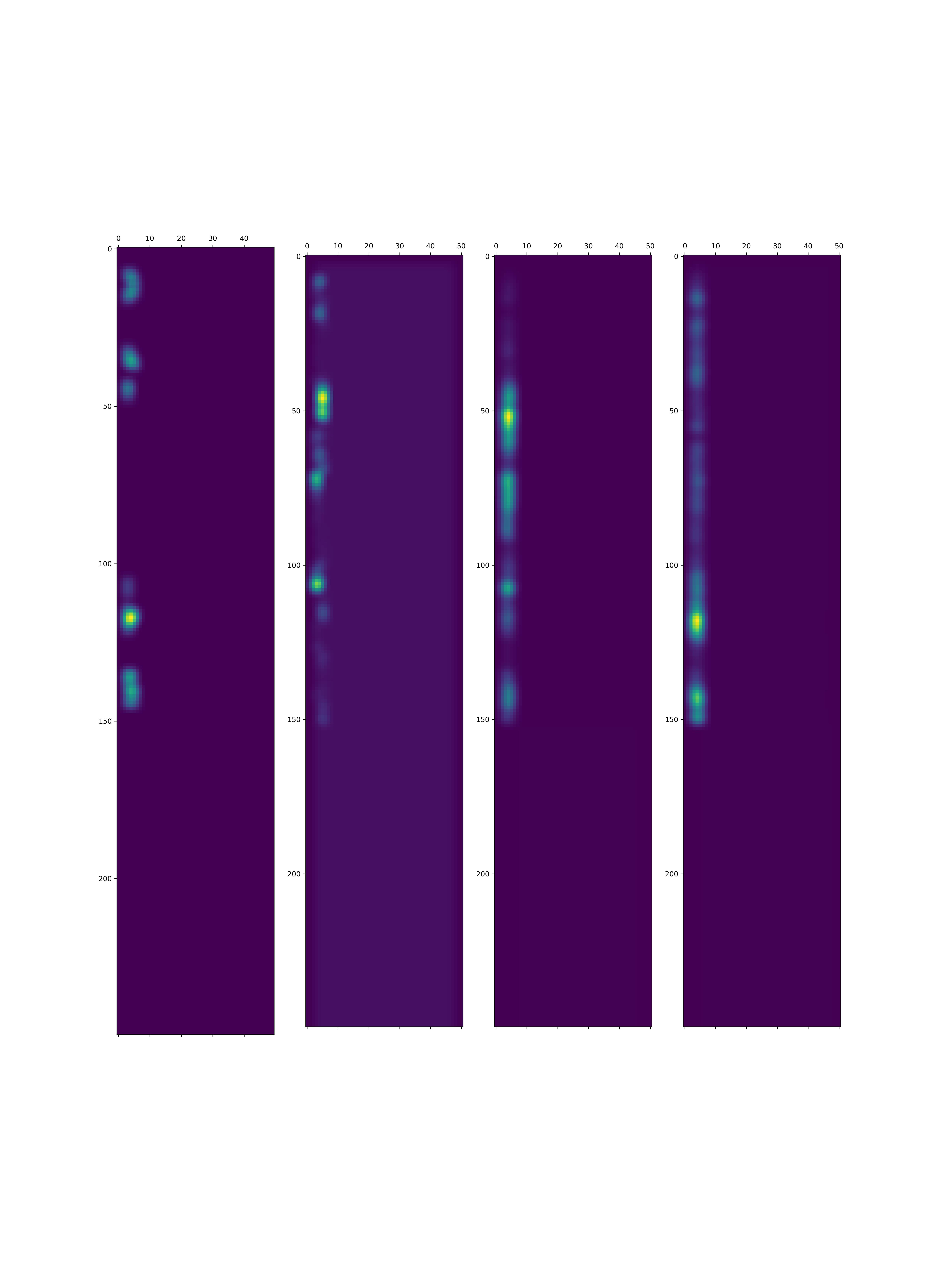}
  \end{minipage}
  \hfill
  \begin{minipage}{0.3\textwidth}
    \includegraphics[width=\textwidth]{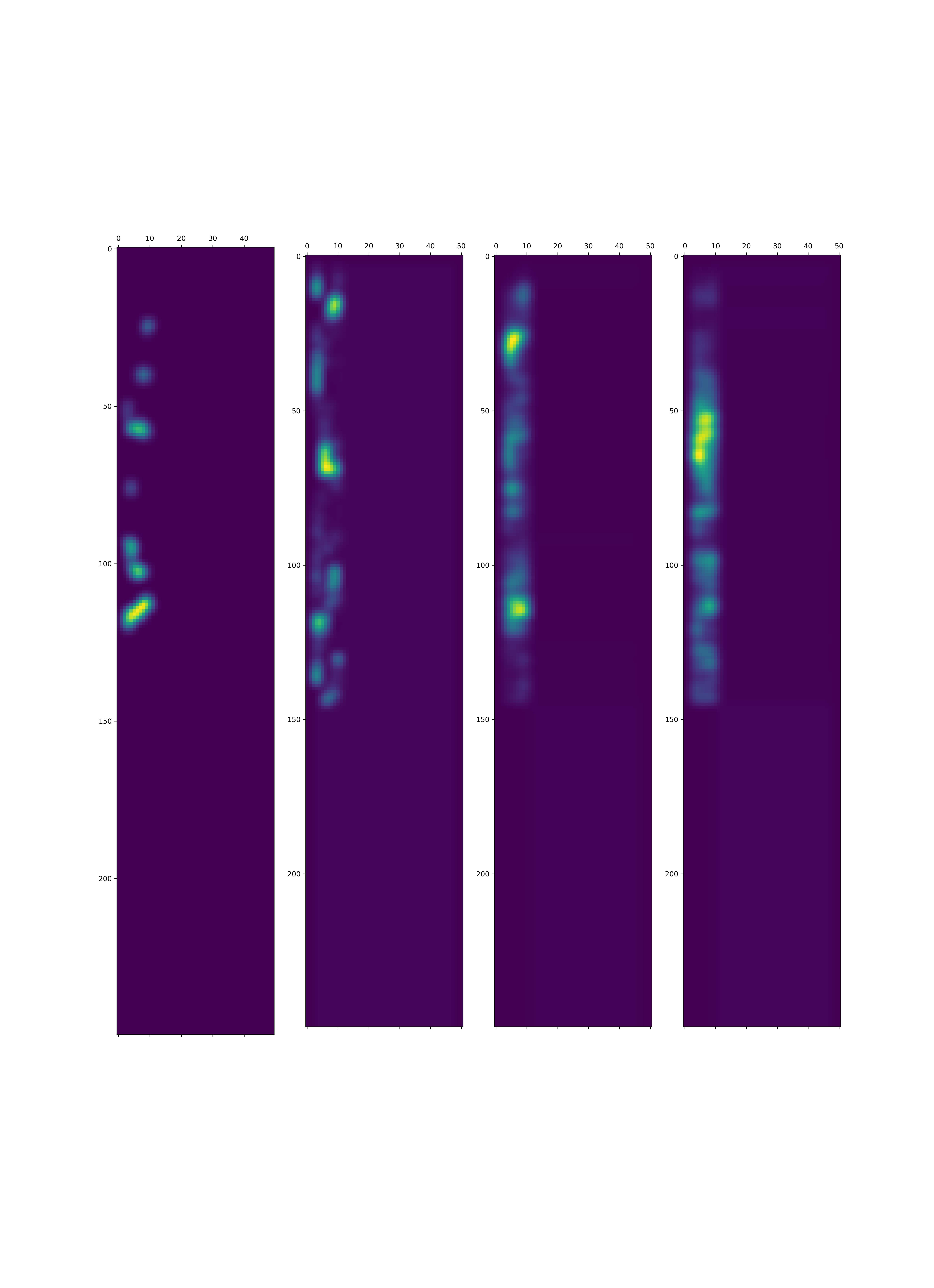}
  \end{minipage}
  \hfill
  \begin{minipage}{0.3\textwidth}
    \includegraphics[width=\textwidth]{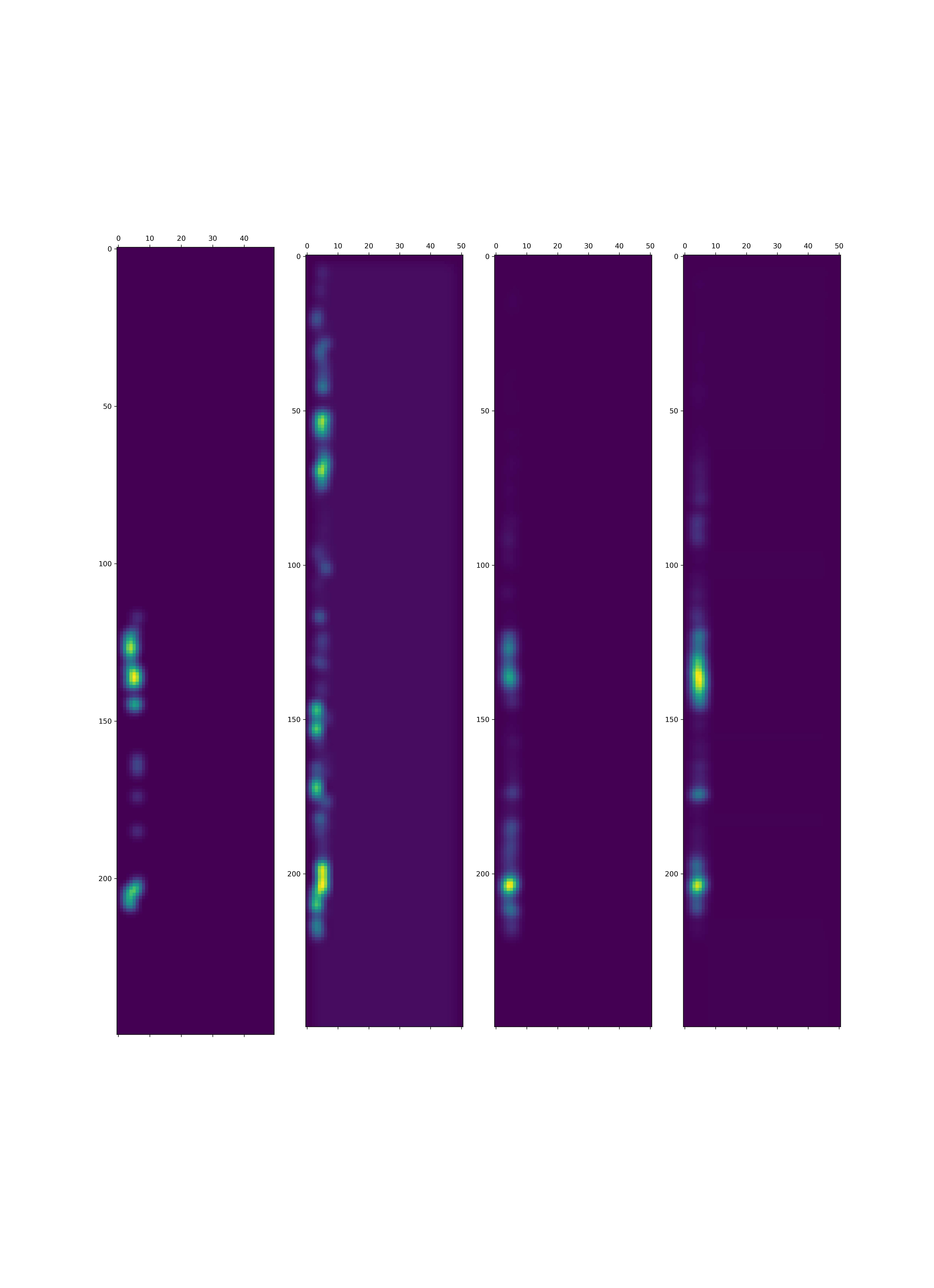}
    
  \end{minipage}
  
  \caption{More predicted examples.}
  \label{fig:example}
\end{figure}
\begin{figure}[htbp]
  \centering
  \begin{minipage}{0.3\textwidth}
    \includegraphics[width=\textwidth]{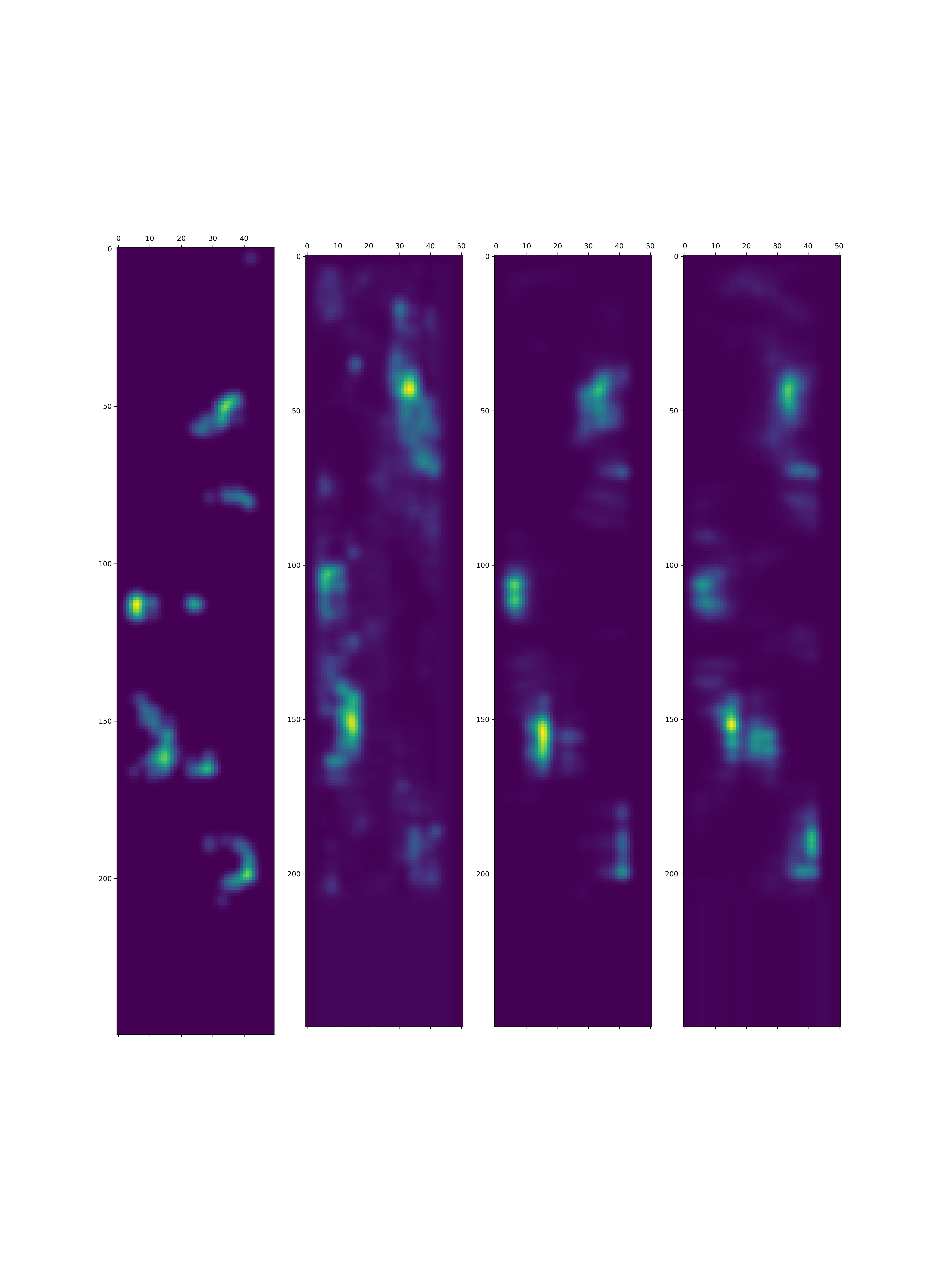}
  \end{minipage}
  \hfill
  \begin{minipage}{0.3\textwidth}
    \includegraphics[width=\textwidth]{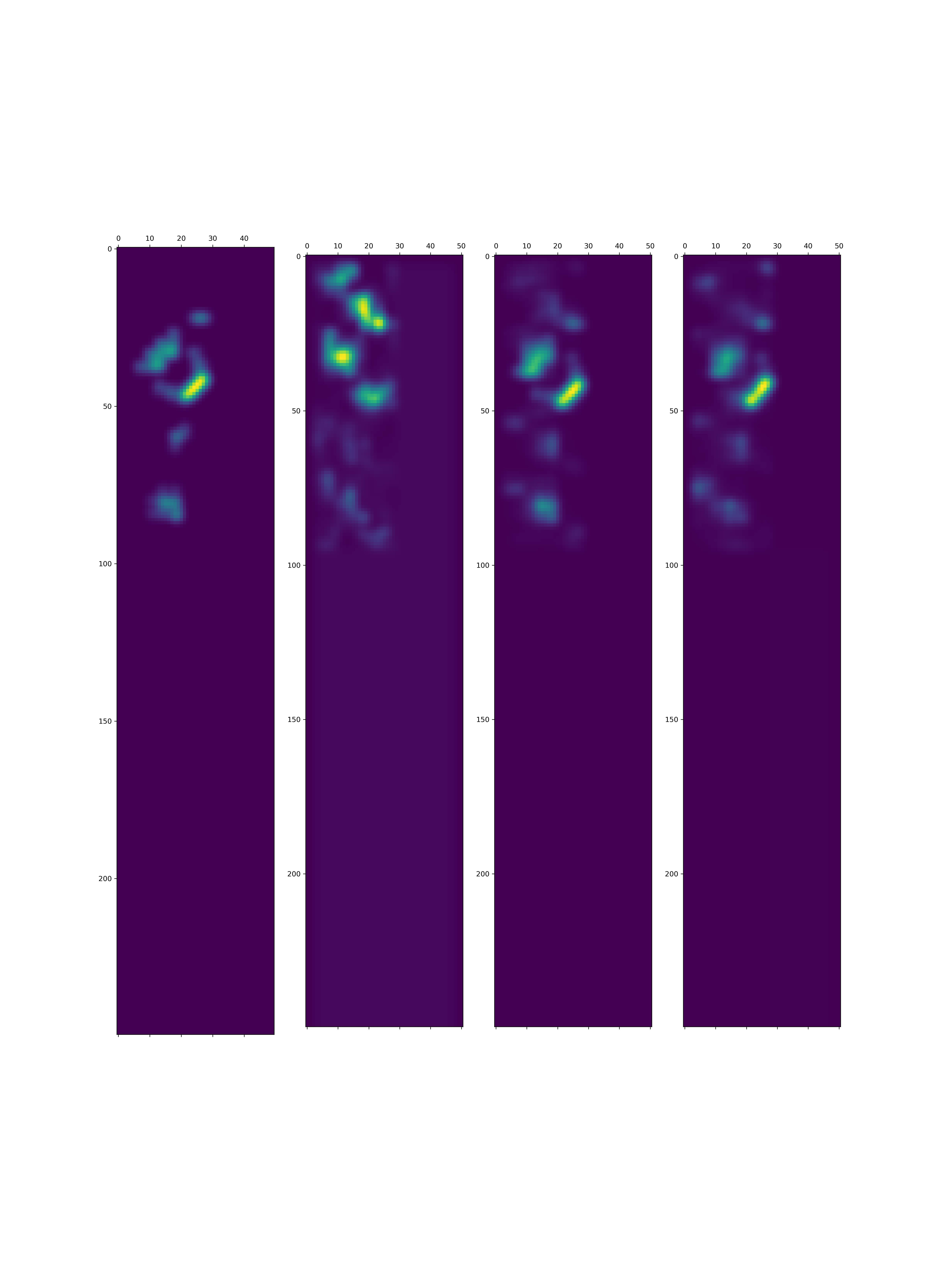}
  \end{minipage}
  \hfill
  \begin{minipage}{0.3\textwidth}
    \includegraphics[width=\textwidth]{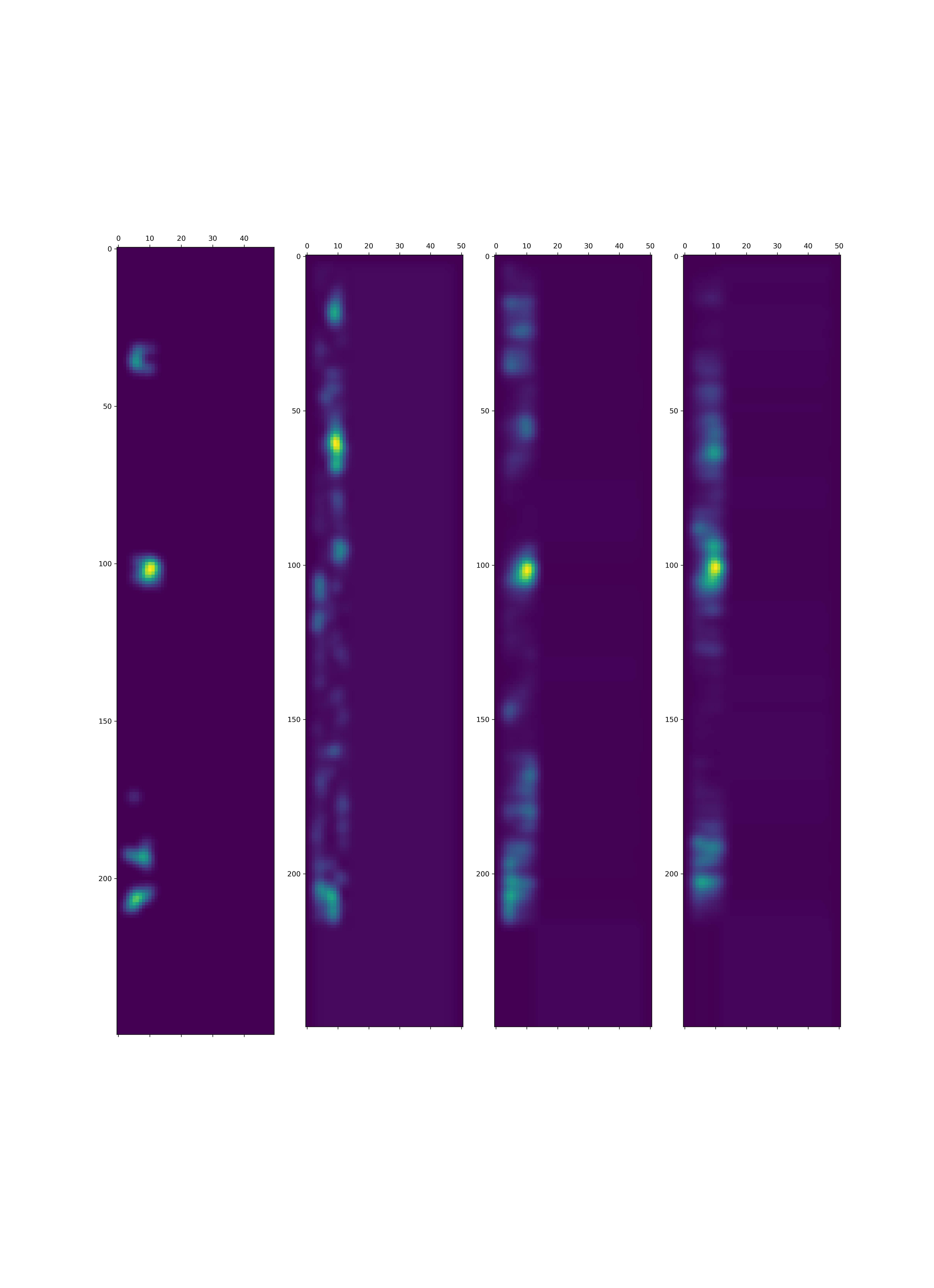}
    
  \end{minipage}
  
  \caption{More predicted examples.}
  \label{fig:example}
\end{figure}
\begin{figure}[htbp]
    \centering
    \includegraphics[width=0.5\textwidth]{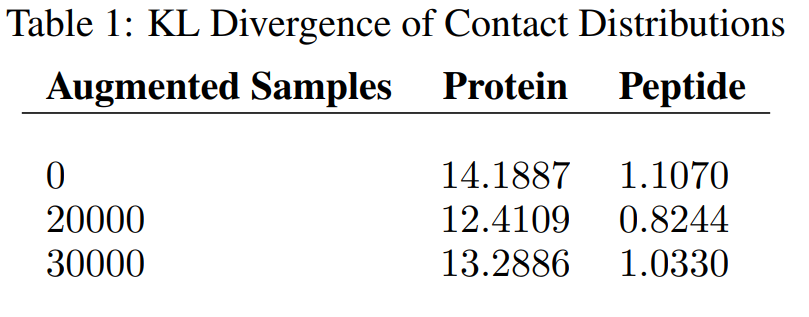}

    \label{fig:your_label}
\end{figure}
\end{document}